\documentstyle[12pt,preprint]{aastex}  

\begin{document}

\title{The Albedos, Sizes, Colors and Satellites of Dwarf Planets Compared with Newly Measured Dwarf Planet 2013 FY27}
\author{Scott S. Sheppard\altaffilmark{1}, Yanga R. Fernandez\altaffilmark{2} and Arielle Moullet\altaffilmark{3}}

\altaffiltext{1}{Department of Terrestrial Magnetism, Carnegie Institution for Science, 5241 Broad Branch Rd. NW, Washington, DC 20015, USA, ssheppard@carnegiescience.edu}
\altaffiltext{2}{Dept. of Physics, Univ. of Central Florida, Orlando, FL 32816, USA}
\altaffiltext{3}{Universities Space Research Association, SOFIA Science Center, Moffett Field, CA 94035 USA}

\begin{abstract}

2013 FY27 is the ninth intrinsically brightest Trans-Neptunian Object
(TNO).  We observed 2013 FY27 at thermal wavelengths with ALMA and in
the optical with Magellan to determine its size and albedo for the
first time and compare it to other dwarf planets.  The geometric
albedo of 2013 FY27 was found to be $p_V = 0.17_{-0.030}^{+0.045}$,
giving an effective diameter of $D = 765_{-85}^{+80}$ km.  2013 FY27
has a size within the transition region between the largest few TNOs
that have higher albedos and higher densities than smaller TNOs.  No
significant short-term optical light curve was found, with variations
less than $0.06\pm0.02$ mags over hours and days.  The Sloan optical
colors of 2013 FY27 are $g-r=0.76\pm0.02$ and $r-i=0.31\pm0.03$ mags,
which is a moderately red color.  This color is different than the
neutral or ultra-red colors found for the ten largest TNOs, making
2013 FY27 one of the largest known moderately red TNOs, which only
start to be seen, and in abundance, at diameters less than 800 km.
This suggests something physically different might be associated with
TNOs larger than 800 km.  It could be that moderately red surfaces are
older or less ice rich and TNOs larger than 800 km have fresher
surfaces or are able to hold onto more volatile ices.  Its also
possible TNOs larger than 800 km are more fully differentiated, giving
them different surface compositions.  A satellite at $0.17$ arcsec
away and $3.0\pm0.2$ mags fainter than 2013 FY27 was found through
Hubble Space Telescope observations.  Almost all the largest TNOs have
satellites, and the relative small size of 2013 FY27's satellite
suggests it was created through a direct collision, similar to
satellites known around the largest TNOs.  Assuming the satellite has
a similar albedo as the primary, it is about 190 km in diameter,
making the primary $D = 740_{-90}^{+85}$ km.

\end{abstract}


\section{Introduction}

2013 FY27 was discovered on UT March 17, 2013 as part of an ongoing
deep and wide survey for extreme Trans-Neptunian objects (TNOs) by
Sheppard and Trujillo (2016).  The orbit of 2013 FY27, with a
semi-major axis near 59 au, eccentricity of 0.39 and inclination of 33
degs, makes it a typical scattered disk object.  With a perihelion
that comes inside of 36 au, 2013 FY27 can come relatively close to
Neptune and have significant gravitational interactions with the
planet.  The high eccentricity of 2013 FY27's orbit means it
experiences large surface temperature variations of some 16 to 22
Kelvin between aphelion and perihelion.

Currently 2013 FY27 is the ninth intrinsically brightest TNO and one
of the most distant at around 80 au, which is near its aphelion.  2013
FY27 is the intrinsically brightest known TNO that has not had its
thermal emission measured (Figure~\ref{fig:KBOalbedo}).  The other
intrinsically brightest 15 or so TNOs have been observed for their
thermal emission by either Spitzer, Herschel or ALMA (Stansberry et
al. 2008; Lellouch et al. 2017; Gerdes et al. 2017; Kovalenko et
al. 2017; Santos-Sanz et al. 2017).  From thermal observations, one
can calculate how much sunlight an object absorbs.  Optical
observations allows one to calculate how much sunlight an object
scatters or reflects.  These two measurements can then be used to
solved for the two unknowns of the size and albedo of an object, using
the radiometric method (Lebofsky et al. 1989; Harris 1998; Fernandez
et al. 2013).

Understanding the size and albedo of 2013 FY27 is important in order
to put this intrinsically bright object into context of the other
dwarf planets beyond Neptune.  The largest TNOs may have formed or
evolved in a much different manner than the more moderate and smaller
objects.  The intrinsic brightness of 2013 FY27 means it is near the
interesting transition regime between the largest few TNOs, which have
high albedos and high densities, and the more moderate and smaller
TNOs, which have moderate and lower albedos and densities (Brown
2013).  There also seems to be some correlation between the color of a
TNO and its albedo and dynamical classification (Lacerda et al. 2014).
Most of the largest TNOs have known satellites, which one can use to
find the bulk densities of the objects.  The densities of the largest
TNOs seem to be much higher ($\sim 2-3$ $g/cm^{3}$) than for the
smaller TNOs ($< 1$ $g/cm^{3}$) (Brown 2013; Grundy et al. 2015; Barr
\& Schwamb 2016).  This suggests the larger TNOs are made of more rock
and less ice and have less porosity than the smaller TNOs.  The
intermediately sized large TNOs, like 2013 FY27, are thus key to
understanding where, how and why this transition in albedo and density
occurs.  Through thermal and optical observations, we place 2013 FY27
into the context of the largest TNOs.

\section{Magellan Optical Observations}

We observed 2013 FY27 using the Magellan 6.5 meter telescope at Las
Campanas, in Chile on UT March 8, 9 and 10 and May 3, 2016.  The IMACS
camera was used, which has a pixel scale of 0.20 arcsec per pixel and
field of view of about 0.16 square deg. The geometry of the
observations are shown in Table 1.  All images were bias subtracted
and flat fielded with nightly dithered sky twilight flats.  Images
were obtained guiding at sidereal rates in photometric conditions
using the Sloan g, r or i-band filters.

Most of the Magellan images were obtained in the r-band to look for
short-term variability of 2013 FY27 over minutes, hours and days.  To
determine the optical color of 2013 FY27, we observed 2013 FY27
multiple times on UT March 10, 2016 in the g-band and i-band Sloan
filters as well as the r-band filter.  The filters were rotated
between the r, g and i-bands to prevent any possible short-term
variations from effecting the color measurements and were obtained
twice, separated by a few hours.  We also convert the Sloan g,r,i
colors to the Johnson-Morgan-Cousins BVRI filter system for easier
comparison to some past works using the transformation equations from
Smith et al. (2002): $B=g'+0.47(g'-r') +0.17$;
$V=g'-0.55(g'-r')-0.03$; $V-R=0.59(g'-r')+0.11$;
$R-I=1.00(r'-i')+0.21$.  Sheppard (2010) showed these transformations
from g,r,i colors to BVRI colors are good to within a hundredth of a
magnitude for most TNOs.

All the optical observations were calibrated using the Sloan standard
star fields DLS1359-11 and PG1633+099.  Seeing was between 0.6 and 1.5
arcseconds, with longer exposure times used when the seeing was worse.
The photometry as well as the details of each individual image from
the observations are shown in Table 2.  All photometry was obtained
similar to that described in Sheppard (2007) and used apertures to fit
the variable seeing, ranging from 2.4 in the best seeing to 4
arcseconds in the worse seeing.  The background counts for each image
were subtracted off the photometry of the object using an aperture
annulus that was larger than that used for the object's photometry.

\subsection{r-band Optical Light Curve}

\begin{figure}
\epsscale{0.4}
\centerline{\includegraphics[angle=90,totalheight=0.4\textheight]{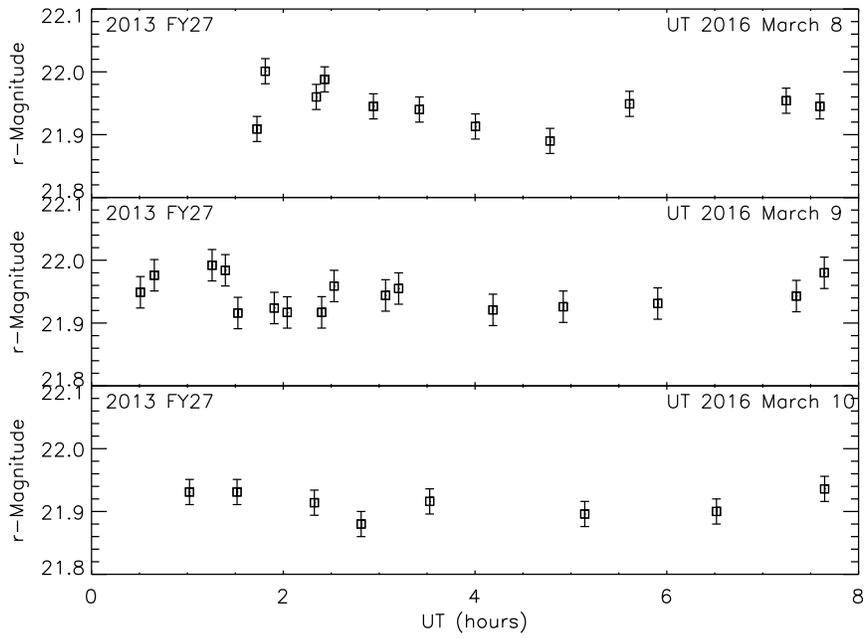}}
\caption{r-band photometry over several hours and days for 2013 FY27
  from the Magellan telescope on UT March 8, 9 and 10, 2016.  No
  obvious rotational light curve is seen over this time, with any
  variations less than about $0.06\pm0.02$.}
\label{fig:multify27} 
\end{figure}

We observed 2013 FY27 in the r-band at Magellan over minutes, hours
and days in early March 2016 to look for any short-term variations
caused by rotation of the object.  The observational results are shown
in Table 2.  No clear variability was seen, with variability being
less than $0.06\pm 0.02$ over the 3 days of observations from March 8
to March 10, 2016 (Figure~\ref{fig:multify27}).  This suggests 2013
FY27 either has a very low amplitude rotational variability, or that
its rotation is longer than a few days time.  As 2013 FY27 is a large
TNO, it should be mostly spherical in shape due to the strength of its
own gravity.  Any short-term variations would likely be from albedo
differences on its surface, which we see no strong evidence for.
There is also the less likely possibility that we are observing 2013
FY27 pole-on, and thus would see no rotational short-term variability.

\subsection{Sloan Optical Colors}

\begin{figure}
\epsscale{0.4}
\centerline{\includegraphics[angle=90,totalheight=0.4\textheight]{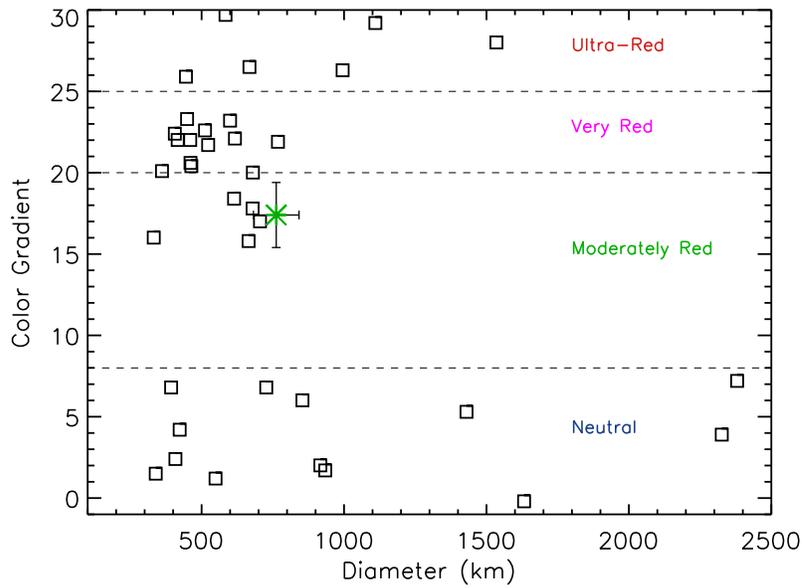}}
\caption{The colors of the largest TNOs.  It is apparent that the
  largest ten TNOs have either very neutral or ultra-red colors.  Only
  objects with diameters below about 800 km start to show moderately
  red colors.  Since neutral and ultra-red material are associated
  with volatiles and organics, this might suggest objects greater than
  800 km have modified or resurfaced their surfaces in some way
  compared to smaller objects, possibly from differentiation or
  cryovolcanism.  2013 FY27 is one of the largest known moderately red
  TNOs shown by the green asterisk.  Surprisingly, there are no
  moderately large or larger TNOs with color gradients between about 8
  and 15 while the largest do not have color gradients between about 8
  and 25.  Error bars shown for 2013 FY27 are similar to the error
  bars of the other TNOs which have been removed for clarity.}
  \label{fig:KBOcolorsDwarf} 
\end{figure}

Using the Sloan $g$, $r$ and $i$-band observations from Magellan
described above, we find the colors of 2013 FY27 are $g-r=0.76\pm0.02$
and $r-i=0.31\pm0.03$ giving $g-i=1.07\pm0.04$ mags (Table 3).  Since
the majority of TNOs have nearly linear color slopes at visible
wavelengths, the basic color of a TNO can be reported as its spectral
gradient (see Doressoundiram et al. 2008 and Sheppard 2010).  The
spectral gradient is the amount of reddening per 100 nanometers and
can be found through

\begin{equation}
  S(\lambda_{2} > \lambda_{1}) = (F_{2,V} - F_{1,V}) /(\lambda_{2} - \lambda_{1})
\end{equation}

where $\lambda_{1}$ and $\lambda_{2}$ are the midpoint wavelengths for
the filters used and $F_{1,V}$ and $F_{2,V}$ are the flux in the
filters normalized to the V-band.  We use the $g$ and $i$ filters to
determine $S$ for 2013 FY27 as these filters have well separated
central wavelengths.  For 2013 FY27, we find the optical color
spectral slope is $S=17\pm 2$.  This color is moderately red when
compared to other TNO colors taken from Hainaut et al. (2012) and the
updated Minor Bodies in the Outer Solar System (MBOSS) data base
(Figure~\ref{fig:KBOcolorsDwarf}).

It has been known since the first Kuiper Belt objects were discovered
that they exhibit a very wide range of colors with possible color
groupings (Luu \& Jewitt 1996; Barucci et al. 2005).  Ultra-red
material is generally considered to have a large red spectral gradient
of $S > 25$ (Jewitt 2002; Sheppard 2010,2012).  In Sheppard (2012)
there is a noticeable gap in objects from the Cold Classical belt with
colors just below $S \sim 20$, so we define $20 < S < 25$ as very
red objects.  Objects with spectral gradients between about 8 and 20
are considered moderately red while objects with color below this are
considered neutral to blue in color.

It is apparent in Figure~\ref{fig:KBOcolorsDwarf} that the ten largest
TNOs only show extremes in colors, being either very neutral or
ultra-red.  It is not until objects smaller than about 800 km in
diameter start to show very red and moderately red colors.  This is
interesting as the extreme colors are usually associated with organics
and freshly exposed ices such as Methane, water ice and methanol
(Brown et al. 2012; Dalle Ore et al. 2015; Fraser et al. 2015).  This
suggests the largest TNOs are continuing or have more recently
modified their surfaces compared to the TNOs smaller than 800 km.
Moderately red colors might be more expected on objects that have old
surfaces as micrometeorite bombardment and irradiation would be
expected to dull any exposed ices over time (Grundy 2009).  The
surfaces of the largest TNOs might be fresher because they can more
easily hold on to volatile ices and could further have differentiated
more completely and even have cryovolcanism occurring in recent times.
2013 FY27 is one of the largest known moderately red objects.

\subsection{Optical Phase Curve}

The optical apparent magnitude of a TNO depends on its radius ($r$),
distance from the Sun ($R$), distance from Earth ($\Delta$), albedo
($p$) and phase angle ($\alpha$).  The apparent optical magnitude
can be calculated as,

\begin{equation}
m_{filter}=m_{\odot}-2.5\mbox{log}\left[p*r^{2}\phi (\alpha )/(2.25\times 10^{16}R^{2}\Delta^{2})\right]  \label{eq:appmag}
\end{equation}

where $m_{\odot}$ is the apparent magnitude of the Sun in the filter
being used and the linear phase function $\phi (\alpha)$ can be represented as

\begin{equation} 
\phi(\alpha) = 10^{-0.4\beta \alpha}
\label{eq:phangle}
\end{equation}

where $\alpha$ is the phase angle in degrees and $\beta$ is the linear
phase coefficient in magnitudes per degree.  At opposition, $\alpha=0$
deg, and thus $\phi (0)=1$.

\begin{figure}
\epsscale{0.4}
\centerline{\includegraphics[angle=90,totalheight=0.4\textheight]{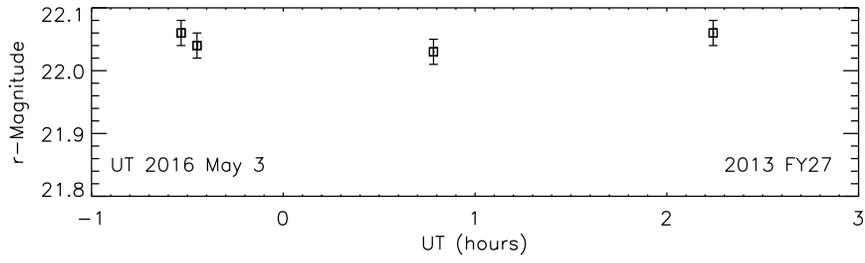}}
\caption{r-band photometry over a few hours for 2013 FY27 from the
  Magellan telescope on UT May 3, 2016.  Again, no obvious short-term
  variations are seen.  Combining the May 2016 observations with the
  March 2016 observations allows for the determination of the phase
  curve for 2013 FY27.}
\label{fig:multify27may} 
\end{figure}

We further observed 2013 FY27 at Magellan on May 3, 2016 over several
hours to determine its phase coefficient
(Figure~\ref{fig:multify27may}).  Since no significant short-term
light curve was found in March, any long-term variations in brightness
are likely attributed to the different phase angles observed for 2013
FY27.  The March 2016 observations were obtained at a phase angle near
0.18 degrees while the May observations had the object further from
opposition, near 0.62 degrees (Table 1).  Again, 2013 FY27 showed no
significant variations over the nearly three hours of observations on
May 3, 2016.  The average r-band magnitude in March was $21.92\pm
0.02$ mags and in May $22.04\pm 0.02$ mags.  2013 FY27 should have
been about 0.01 mags fainter in May because it was slightly further
from the Earth at that time.  Thus 2013 FY27 was about 0.11 mag
fainter in May when accounting for differences in its distance from
the Sun and Earth.  We attribute the 0.11 mag difference from 2013
FY27 being 0.44 degrees further away from true opposition in May than
the March observations.

\begin{figure}
\epsscale{0.4}
\centerline{\includegraphics[angle=90,totalheight=0.4\textheight]{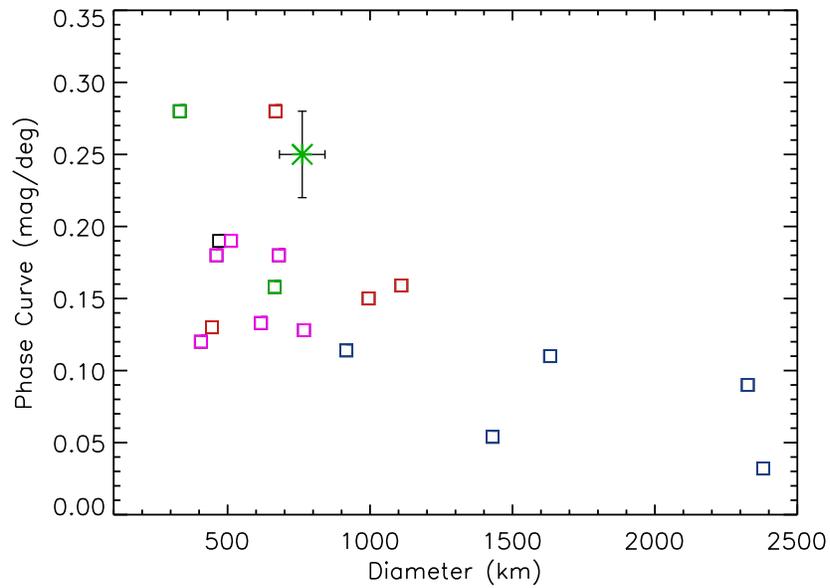}}
\caption{The phase coefficients, $\beta$, for the largest TNOs.  The
  largest few TNOs, all with neutral colors and high albedos, have
  generally lower phase coefficients.  Most other TNOs have phase
  coefficients between about 0.12 and 0.20 mags per deg.  2013 FY27 is
  shown by the asterisk and appears to have a higher than normal phase
  coefficient.  Errors have been removed for clarity, but are
  generally a few hundredths of a mag per deg.}
\label{fig:KBOphaseDwarf} 
\end{figure}

\begin{figure}
\epsscale{0.4}
\centerline{\includegraphics[angle=90,totalheight=0.4\textheight]{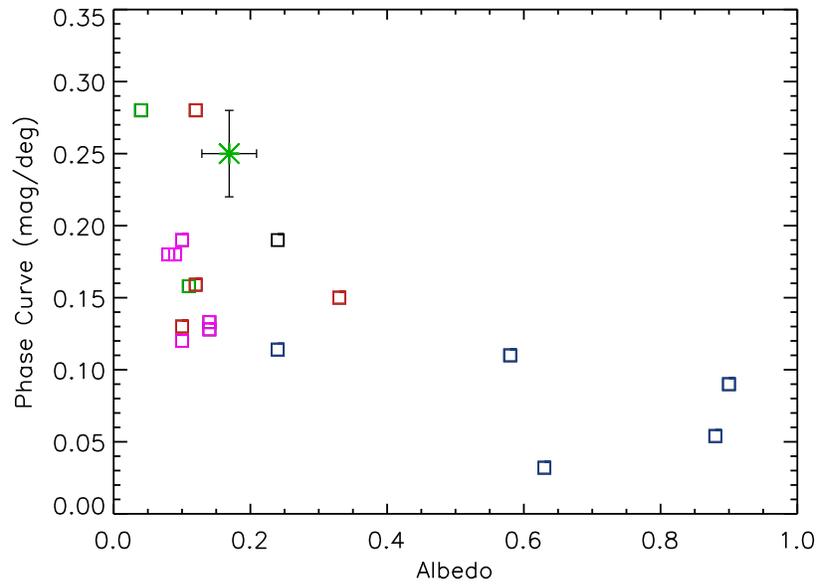}}
\caption{Same as Figure~\ref{fig:KBOphaseDwarf}, but comparing albedo
  instead of diameter to the phase coefficient.  It is clear the
  few very high albedo objects are all neutral in color and have the lowest
  phase coefficients.}
\label{fig:KBOphase2Dwarf} 
\end{figure}

From the above March and May r-band observations, we find a phase
coefficient of $\beta = 0.25\pm 0.03$ mags per deg for 2013 FY27.  We
show in Figure~\ref{fig:KBOphaseDwarf} the phase coefficients for the
largest TNOs using data from Buie et al. (1997); Sheppard and Jewitt
(2002),(2003); Rabinowitz et al. (2007); Sheppard (2007); Benecchi \&
Sheppard (2013).  We confirm the finding in Sheppard (2007) that the
largest few TNOs, which are all neutral in color with high albedos,
show the lowest phase coefficients (Figure~\ref{fig:KBOphase2Dwarf}).
The lower phase coefficient is likely from back scattering off a high
albedo surface.  2013 FY27 has a higher than typical TNO phase
coefficient, which are mostly in the 0.10 to 0.20 mag per degree
range.  This higher phase coefficient could signify different grain
properties on 2013 FY27's surface compared to the typical TNO or might
be a sign of a very long and significant rotational light curve that
would not be seen in a 3 day or less period of time.

For main belt asteroids, it appears the higher the phase coefficient,
the lower the surface albedo (Belskaya and Shevchenko 2000).  But main
belt asteroids likely have different compositions and usually have
lower phase coefficients than the TNOs.  In addition, most asteroid
phase curves are based over a much larger range of phase angles, in
which the TNOs are not able to be viewed as they typically stay below
about 2 degrees as seen from Earth.  The opposition surge, when an
object gets much brighter at very low phase angles from back-scattering
effects, starts around 0.1 to 0.2 degrees (Belskaya et al. 2008).
Thus the 2013 FY27 observations are just outside of the start of the
opposition surge and likely not strongly affected by the opposition
surge.

\subsection{Optical Absolute Magnitude}

Using the linear phase coefficient found above, we can calculate the
reduced magnitude as

\begin{equation}
m_{R}(1,1,0) = m_{R} - 5\mbox{log}(R\Delta).
\end{equation}
    
The reduced magnitude of a solar system object is the magnitude it
would have if it were observed simultaneously at a geocentric and
heliocentric distance of 1 au with a phase angle of 0 deg.  Thus the
reduced magnitude gives you the brightness of an object independent of
observing geometry.  The reduced magnitude brightness is for the most
part only dependent on the size and albedo of an object.

Using the above equation, we find the reduced magnitude of 2013 FY27
at $\alpha = 0.18$ deg as $2.90\pm 0.02$ mags in the r-band during the
March 2016 observations.  For the May 2016 observations, we find the
reduced magnitude is $3.01\pm 0.02$ mags at $\alpha = 0.62$ deg
because the object is further from opposition and thus fainter from
showing a less illuminated face towards Earth.

To calculate the reduced magnitude of 2013 FY27 when at zero degrees
phase angle, we use the linear phase coefficient found above of $\beta
= 0.25$ mags per deg.  So 2013 FY27 should be $0.25$ mags$/$deg
$\times$ $0.18$ deg $= 0.05$ mags brighter when at 0 deg phase angle
compared to 0.18 deg phase angle assuming a linear phase function.
Thus we find $m_{r}(1,1,0)=2.85\pm 0.02$ mags.  From the
transformation equations in Smith et al. (2002), we find the
Johnson-Kron-Cousins R-band reduced magnitude is $m_{R}(1,1,0)=2.59\pm
0.02$ mags.  Using the color found for 2013 FY27 of $V-R=0.56\pm 0.03$
mags, we find $m_{V}(1,1,0)=3.15\pm 0.03$ mags, which we take as the
absolute magnitude $H$ of 2013 FY27.  This is slightly fainter than
the current value used at the minor planet center for 2013 FY27, which
is $H=3.0$ mags.  This slight difference in the absolute magnitudes is
to be expected as Sheppard (2007) found the Minor Planet Center is
routinely off by several tenths of magnitudes from well measured
calculated values of the absolute magnitude, likely because the Minor
Planet Center uses all photometry from multiple sources and filters,
some of which have low signal to noise and thus large uncertainties.
Though the absolute magnitude $H$ generally uses a curved phase
function as defined in Bowell et al. (1989), Sheppard (2007) found the
reduced magnitude for TNOs using a linear phase function with $\beta$
is similar to within a few hundredths of the absolute magnitude phase
function used in Bowell et al. (1989).

An important optical reference is to determine the optical brightness
of 2013 FY27 at the time of the ALMA observations in late December
2017 and early January 2018.  The phase angle and distance of 2013
FY27 at the time of the ALMA observations is very similar to the May
3, 2016 observations from Magellan.  Thus we should be able to use the
May 3, 2016 photometry as the base for the optical photometry during
the ALMA observations.  At this time $m_{r}=22.04\pm0.02$,
$m_{g}=22.80\pm0.03$ and $m_{i}=21.73\pm0.03$ mags, giving from color
transformations using Smith et al. (2002) $m_{R}=21.79\pm0.02$,
$m_{V}=22.35\pm0.03$ and $m_{I}=21.27\pm0.03$ mags.

\section{ALMA Observations}

The Atacama Large Millimeter Array (ALMA) observations were taken on
UT 29 and 30 December 2017 and 04 January 2018 in ALMA Cycle 5 under
Project 2017.1.01662.S (Table 1). At that time, the array was
configured with 46 antennas and a maximal distance between antennas
(maximum baseline) of 2500~m. We used Band 3 in the standard continuum
setup centered at 97.5 GHz, with a total encompassed bandwidth of 8
GHz.  For such a cold object, Band 3 provides the best sensitivity in
the continuum among all the ALMA bands making it the best to use as a
first thermal detection attempt.  The ALMA observations consisted of
three sets of 66 minutes observations, one each on 29 and 30 December
2017 and another on 04 January 2018.  The combined observations
totaled of 3.3 hours of ALMA time, with 2.2 hours of that integrating
on the source. The rest of the observation time was spent on quasars
used as bandpass, flux and gain calibrators.

Interferometric measurements consist in visibilities, which are
complex numbers corresponding to signal cross-correlations between
each pair of antennas. For each of the observations, visibilities were
calibrated using the ALMA calibration pipeline in the Common Astronomy
Software Applications package (CASA, McMullin et al. 2007), to correct
for the spectral response and temporal gain variations of the
instrument. The absolute flux scale was assessed using the well
monitored quasar J1127-1857 as a reference.

\begin{figure}
\epsscale{0.4}
\centerline{\includegraphics[angle=0,totalheight=0.6\textheight]{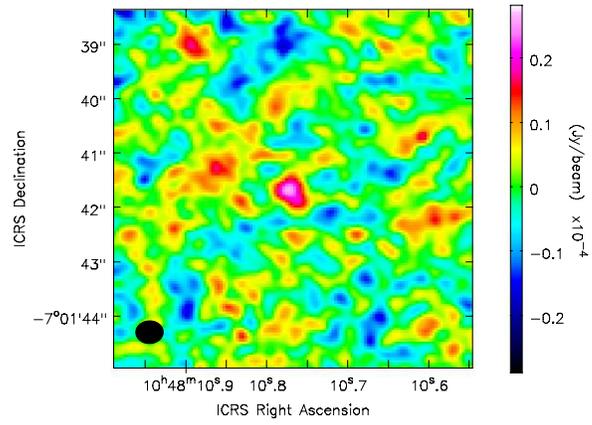}}
\caption{Stacked ALMA image showing the detection of 2013 FY27, which
  appeared at the center of the image as expected.  The ALMA beam size
  is shown by the black oval in the lower left.}
\label{fig:ALMAimage} 
\end{figure}

The three sets of observations were then combined, spectrally averaged
and stacked into one visibility set. To obtain a continuum image,
inverse Fourier transform and deconvolution were applied to the
combined visibilities set. Since this experiment is a point-source
detection which does not depend on spatial resolution, natural
weighting was used to maximize sensitivity (at the expense of beam
size), giving the final image an rms of 5.5 microJy. The synthesized
beam was $0.50 \times 0.42$ arcseconds in the natural-weighted image.
2013 FY27 was found very near the center of the image as expected, and
was 25 microJy bright (Figure~\ref{fig:ALMAimage}).  This gives the
ALMA detection of 2013 FY27 a Signal-to-Noise of about 4.5. To assess
the quality of the flux calibration, we compared the results given by
the calibration pipeline using J1127-1857 as a reference to the
expected flux value for the phase calibrator J1058+0133, which is a
relatively bright and well monitored quasar. The good match between
the retrieved and expected flux for J1058+0133 allow us to determine
that the absolute flux calibration is accurate at the 3 to 4 percent
level.

\section{Size and Albedo of 2013 FY27}

To first order, the thermal flux from an object is proportional to
$D^2(1-p_Vq)$, where $D$ is the effective diameter, $p_V$ is the
geometric albedo in V-band, and $q$ is the phase integral.  The
reflected sunlight in the V-band from the same object is proportional
to $D^2p_V$.  As shown in Brown \& Butler (2017), $q$ for TNOs can be
approximated by $0.336p_V + 0.479$.  Thus we have two equations that
can be solved for the two unknowns of diameter and albedo ($D$ and
$p_V$) (see e.g. Fernandez et al. 2013).

The most common method to determine the size and albedo of an object
with single-epoch thermal photometry, as here with 2013 FY27, is to
use the Near-Earth Asteroid Thermal Model (NEATM) of Harris (1998).
Some basic assumptions used in NEATM are that (i) the object is
spherical; (ii) the phase darkening of the thermal emission is
entirely dependent on how much of the lit-up hemisphere of the object
is facing Earth, and (iii) the basic dayside surface temperature falls
off from the subsolar point as $\cos^{1/4}\theta$, where $\theta$ is
the local zenith angle, and there is no thermal emission from the
nightside.

These NEATM assumptions are reasonable for 2013 FY27.  2013 FY27 is
probably close to spherical as it is likely large and there is little
if any apparent shape-induced rotational modulation of our optical
photometry (section 2.1).  Our observations were obtained at very
small phase angles and thus phase darkening uncertainties should be
minimal.  Regarding the temperature map, the NEATM is appropriate for
a ``slow rotator," i.e. an object whose thermal inertia is
sufficiently low, and rotation period sufficiently long, that the
object has no thermal memory. In NEATM, deviation from a zero-thermal
memory situation is encapsulated in the beaming parameter, $\eta$,
which will account for the cooler temperatures expected for an object
with some thermal memory. A value of $\eta=1$ corresponds to the zero
thermal memory situation, and a value of $\eta>1$ applies to cases
with thermal memory, with the higher the value, the more memory (such
as the case of a quick rotator or a large thermal inertia). For our
analysis of 2013 FY27, since we have measurements at only one thermal
wavelength, we cannot independently determine what the beaming
parameter should be. However we can use other TNO measurements to
examine what would be appropriate to assume. Lellouch et al. (2013,
2017), in a study of the thermal emission from several TNOs as
observed with ALMA, Herschel, and Spitzer, found that the thermal
inertias of these bodies are quite low, at least an order of magnitude
below that of the Moon.  It is likely safe to assume that 2013 FY27 is
similar, and so should be close to acting like a slow rotator in the
NEATM model for reasonable values of its rotation period.  Lellouch et
al. (2017) advise using a value of $\eta=1.175$ for TNOs observed with
ALMA, so we adopt that value here.  The other assumptions that go into
NEATM are the emissivity and the visible-wavelength phase law. For the
former, we assume a value of $0.70$, as suggested by Lellouch et
al. (2017). For the latter, we apply our results from section 2.3,
where $\beta = 0.25$ mag/deg.

\begin{figure}
\epsscale{0.4}
\centerline{\includegraphics[angle=0,totalheight=0.25\textheight]{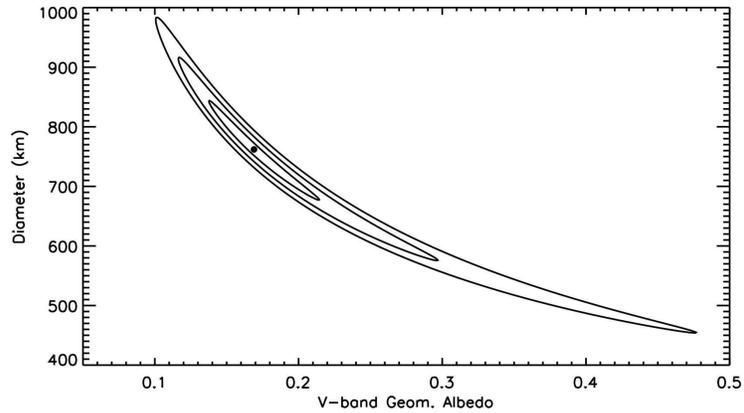}}
\caption{Contour plot showing the goodness of fit of NEATM to our
  millimeter and visible photometry that results in estimates of 2013
  FY27's diameter and geometric albedo.  The center dot is the
  location of the best fit, and the contours extending out from that
  point represent unreduced $\chi^2$ values of 1, 2, and 3.  We take a
  good fit to be where $\chi^2\le 1$, giving an effective diameter of
  $D = 765_{-85}^{+80}$ km and albedo of $p_V =
  0.17_{-0.030}^{+0.045}$ for the 2013 FY27 system.}
  \label{fig:YanAlbedoSize} 
\end{figure}

The results of our application of the NEATM to the 2013 FY27
photometry is shown in Figure~\ref{fig:YanAlbedoSize}. We used a flux
density from thermal emission of $25.0\pm5.5$ $\mu$Jy at a wavelength
of 3.075 mm, and a V-band magnitude of $22.35\pm0.03$ to fit $D$ and
$p_V$. The thermal emission error bar incorporates the $\sim$4\%
uncertainty in the absolute calibration.  Since there are zero degrees
of freedom (using two measurements to fit two parameters), we cannot
formerly calculate a reduced-$\chi^2$; instead we use $\chi^2$
itself. The contour plot in Figure~\ref{fig:YanAlbedoSize} shows
contours of $\chi^2=1$, 2, and 3, and we report an error bar on the
best-fitting parameters to correspond to the $\chi^2=1$ level. We find
that the best fit for 2013 FY27 is an effective diameter of $D =
765_{-85}^{+80}$ km and $p_V = 0.17_{-0.030}^{+0.045}$. The two error
bars are correlated, as the elongated contours show.  The errors on
$D$ and $p_V$ we report here are from the uncertainty in the
photometry including the absolute calibration of the standard sources
in both the visible and thermal.  Incorporating the potential
uncertainty in the assumed parameters such as the beaming parameter
and effective slow rotation period is tricky, but could add some five
percent uncertainty to the diameter calculation and 10 percent
uncertainty to the albedo calculation.  As noted in section 5 below, a
satellite some 3 magnitudes fainter than 2013 FY27 was found in HST
data. Assuming the satellite has a similar albedo as the primary, the
satellite is about 190 km in diameter, making the primary slightly
smaller than calculated above at $D = 740_{-90}^{+85}$ km.

\subsection{Discussion of Dwarf Planet Diameters, Albedos and Colors}

\begin{figure}
\epsscale{0.4}
\centerline{\includegraphics[angle=90,totalheight=0.4\textheight]{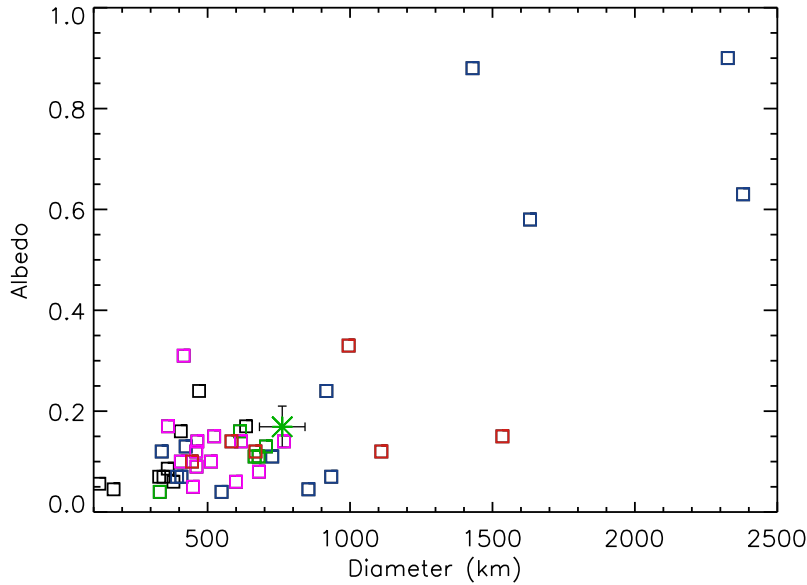}}
\caption{The size versus albedo of the largest TNOs with well
  determined characteristics.  The largest few TNOs have very high
  albedos with the smaller ones much darker.  2013 FY27 is shown by a
  green asterisk with error bars.  2013 FY27 is near the large end of
  objects in the transition region that occurs between the largest and
  smallest TNOs, making 2013 FY27 important in understanding the
  transition region.  Error bars on other TNOs have been removed for
  clarity but are generally similar or smaller than 2013 FY27's error
  bars.  Colors of symbols indicate the visible colors of the
  objects.  It is apparent the largest TNOs are bimodal in albedo and
  color space. The ten largest TNOs are either neutral to blue in
  color (blue symbols) with high albedo or ultra-red in color (red
  symbols) with moderate albedos.  There are no very large TNOs with
  moderately red (green symbols) or very red (magenta symbols) colors.
  2013 FY27 is one of the largest moderately red colored TNOs known.
  The transition from moderately red dominated colors to only neutral
  and ultra-red colors begins around 800 km in diameter.  The albedo
  dispersion also seems to increase near this size.  This might
  indicate TNOs larger than 800 km have altered their surfaces
  compared to the smaller objects.}
\label{fig:KBOalbedo} 
\end{figure}

An albedo of about 17\% is on the high end for a moderately red
object but consistent with the other moderately red and moderately
sized dwarf planet TNOs (Figure~\ref{fig:KBOalbedo}).  2013 FY27
is one of the largest, if not the largest, moderately red TNO.  All
TNOs larger than 2013 FY27, of which there are about ten, have
either ultra-red colors or neutral colors.  The TNOs with similar
sizes and smaller than 2013 FY27 (diameters $< 800$ km) start to
show an abundance of moderately red or very red colors, unlike the
largest ten objects, which do not show any of these middle surface
colors and only the extreme surface colors.

This suggests something is or has physically changed the surfaces of
the largest ten TNOs with diameters above 800 km compared to the TNOs
with diameters smaller than 800 km.  This could be because TNOs above
about 800 km have enough self gravity to retain certain ices such as
methane more readily on their surfaces than the smaller TNOs.  Neutral
colored surfaces are generally associated with fresh water ice while
ultra-red surfaces are associated with organics and possibly other
ices such as methanol and methane.  The color differences may be
because TNOs above 800 km retained enough internal heat to remain
active longer or even to this day, to resurface their surfaces from
possible cryovolcanism type events.  TNOs above 800 km might also be
more fully differentiated than TNOs below 800 km, making their two
surfaces types different in composition.  As TNOs below 800 km still
show neutral and ultra-red surfaces as well as very red and moderately
red surfaces, the moderately red surfaces might be a signature of a very
old surface that has been bombarded by high energy photons, particles
and micrometeorites over long periods of time, while the extreme
colors are more of a sign of fresher surfaces from recent collisions
or activity.  Further analysis is required to determine why the
largest TNOs do not show moderately red surfaces, but
Figure~\ref{fig:KBOalbedo} strongly suggests there is some kind of
surface change around 800 km in diameter for TNOs.

\section{Satellite Discovered around 2013 FY27}

\begin{figure}
\epsscale{0.4}
\centerline{\includegraphics[angle=0,totalheight=0.6\textheight]{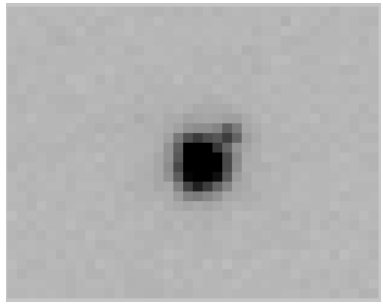}}
\caption{The discovery image of a satellite around 2013 FY27 from the
  Hubble Space Telescope.  The image is in the F350LP filter with the
  satellite $3.0\pm0.2$ mags fainter and about $0.17$ arcseconds with
  a position angle of about 135 degrees with respect to the primary.}
\label{fig:2013FY27moon} 
\end{figure}

We observed 2013 FY27 with the Hubble Space Telescope (HST) on UT
January 15, 2018 to look for possible satellites (HST Program
GO-15248).  Four 545 second images were taken with HST between 01:24
and 02:30 hours UT in the F350LP wide band filter (central wavelength
of 5859 Angstroms) using the WFC3/UVIS instrument with a pixel scale
of $0.04$ arcsec per pixel (MacKenty et al. 2014).  An obvious point
source was detected about $0.17$ arcseconds at a position angle of 135
degs from the primary (Figure~\ref{fig:2013FY27moon}).  The satellite
was seen in all four images and moved along with the motion of the
primary, which was about -1.47 and 0.34 arcseconds an hour in Right
Ascension and Declination, respectively.  This detection was reported
as a satellite of 2013 FY27 to the International Astronomical Union
(see CBET 4537: Sheppard 2018).  No satellite motion relative to the
primary was detected between the first and last (fourth) image from
HST, confirming its association with the primary and likely ruling out
an extremely fast orbital period of a few days or less around the
primary.  The four HST images were aligned with respect to the primary
and coadded to search for additional fainter satellites, but nothing
obvious was detected to about 27.5 mags within a tenth to tens of
arcseconds of the primary.

At the time of the HST observations, 2013 FY27 was 79.48 au from the
Earth.  Thus a 0.17 arcsecond separation means the satellite was
at least 9800 km from the primary.  The newly discovered satellite was
$3.0\pm0.2$ mags fainter than the primary in the optical.  Thus the
diameter of the satellite, if assuming the same albedo as the primary,
would be about 3.9 times smaller than the primary or about 190 km in
diameter.

Additional HST observations of 2013 FY27 and its satellite have been
obtained in May and July of 2018 under HST Program 15460.  A full
analysis of the new HST data will allow the determination of the
satellites orbit and with the albedo reported in this work, a bulk
density of the system can be determined giving insight into the
composition and structure.  The full detailed analysis using all of
the HST observations for 2013 FY27 will be part of a future paper on
the satellite and its orbit around 2013 FY27.  We note that in
Figure~\ref{fig:ALMAimage}, there appears to be an extension of the
ALMA signal to the southwest of the primary.  This is unlikely to be
the satellite of 2013 FY27 as we believe the satellite is nearly
edge-on, and thus should only show up to the southeast and northwest
of the primary.  Obtaining the full orbit of the satellite will allow
us to predict where it should have been during the ALMA observations
and further analysis of the ALMA data and 2013 FY27's satellite is
left for the next paper on the full orbit of the 2013 FY27 system.

\subsection{Discussion of Dwarf Planet Satellites}

\begin{figure}
\epsscale{0.4}
\centerline{\includegraphics[angle=90,totalheight=0.4\textheight]{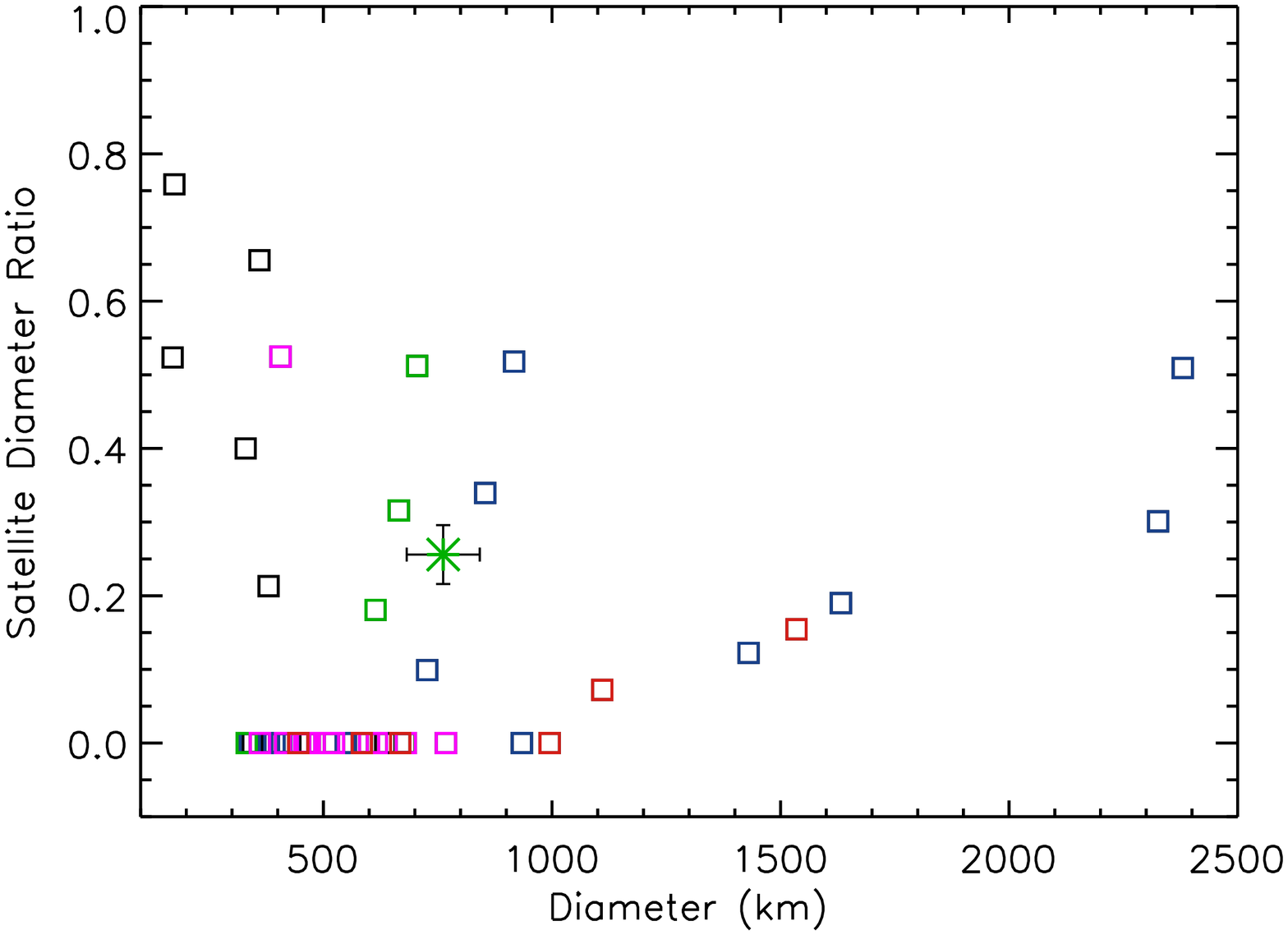}}
\caption{The largest known TNOs and their satellite's diameter to
  primary diameter ratio.  TNOs with a zero size ratio do not have a
  known satellite.  For Pluto and Haumea, which have multiple known
  satellites, the largest known satellite was used to determine the
  satellite to primary diameter ratio.  There appears to be a break
  around 900 km where the satellite to primary diameter ratio becomes
  much smaller for the largest TNOs, with only Pluto being above 0.3.}
\label{fig:KBOsatDwarf} 
\end{figure}

All of the largest known TNOs, those with diameters well over 1000 km,
irrespective of their dynamical class, have known satellites (Pluto,
Eris, Makemake, 2007 OR10, Haumea, and Quaoar) and now 11 of the top
15 largest TNOs have known satellites (Figure~\ref{fig:KBOsatDwarf}).
All of the satellites of these largest objects are significantly
smaller than the primary and have relatively close orbits that are
indicative of collisional formation (Brown et al. 2006; Noll et
al. 2008; Parker et al. 2016; Kiss et al. 2017; Brown \& Butler 2018).
This is remarkable that satellites seem to be the norm and not the
exception for the largest objects.  This appears to suggest the
collisional formation of Earth's moon and the satellites of Mars are
normal outcomes of the planet formation process (Mercury and Venus
likely have no satellites simply because their closeness to the
massive Sun makes tidal interactions too strong for most satellite
orbits around these planets to be stable over the age of the solar
system).

Though we don't know the full orbit of the satellite for 2013 FY27,
its small size relative to the primary and relatively close distance
suggest the satellite was likely created by a direct impact, similar
to the scenarios envisioned for the other known satellites of the
dwarf planets (Canup 2011; McKinnon et al. 2017).  This formation
scenario is quite different than the equal-sized, distant binaries
found mostly in the main Kuiper Belt, which likely did not form
through direct collisions onto the primary (Schlichting and Re'em
2008; Nesvorny et al. 2010; Parker et al. 2011; Sheppard et al. 2012).

Barr and Schwamb (2016) suggest that there are two possible
collisional formation scenarios that could create the close-in
satellites found around all the largest TNOs.  High energy direct
collisions could remove the ice from the primary, leaving a dense
primary and small ice rich satellite.  A glancing indirect collision
would leave the primary volatile rich and thus of low density and
could create larger satellites of similar composition to the primary.
2013 FY27 can be a further test to this theory through determining its
density from knowing the full orbit of the satellite around the
primary as well as looking at the surfaces of the primary and
secondary for similarities or differences.  We note in
Figure~\ref{fig:KBOsatDwarf} there appears to be a break in the ratio
of the satellite size to the primary size for TNOs larger than about
900 km.  TNOs larger than about 900 km have relatively small
satellites compared to the primary (though Pluto somewhat breaks this
trend).  Many of the smaller TNOs appear to have satellites that are
relatively large compared to the primary, with some approaching
equal-sized binaries starting around 400 km in diameter.

\section{Summary}

2013 FY27 is intrinsically the brightest TNO not yet observed for its
basic physical properties.  2013 FY27's absolute magnitude of near 3
mags means it likely has a size that is near the transition region
between the largest few TNOs that show high albedos and high densities
and the smaller TNOs that show low to moderate albedos and densities.
We observed 2013 FY27 over optical and thermal wavelengths at several
telescopes to determine its physical characteristics for the first
time and compare it to the other dwarf planets.

1) The geometric albedo of 2013 FY27 was found to be $p_V =
0.17_{-0.030}^{+0.045}$.  The effective diameter of the 2013 FY27
system is $D =765_{-85}^{+80}$ km.  Assuming then newly discovered
satellite around 2013 FY27 has a similar albedo as the primary, the
satellite is about 190 km in diameter, making the primary slightly
smaller than the effective diameter of the system at $D =
740_{-90}^{+85}$ km.

2) The color of 2013 FY27 was found to be moderately red with a
spectral gradient of $S=17\pm 2$.  This makes 2013 FY27 one of the
largest known moderately red TNOs.  All TNOs larger than about 800 km
in size, of which there are about ten known, only show neutral or
ultra-red surface colors.  This could be because the largest several
TNOs have different or fresher surfaces than TNOs smaller than 800 km
from possible cryovolcanism, differentiation and/or abundant exposed
surface ices.  The TNOs only start to show moderately red colors for
objects less than about 800 km in diameter, and they appear to be very
abundant below this threshold.  As there is also neutral and ultra-red
TNOs below 800 km, it suggests the moderately red color might be an
indication of an old surface while the more neutral and ultra-red
colors in the smaller TNOs could be fresh surfaces exposed from recent
impacts or other processes.

3) A satellite of 2013 FY27 was found in Hubble Space Telescope
observations of 2013 FY27.  It is some $3.0\pm0.2$ mags fainter and
was $0.17$ arcsec away from the primary at discovery.  The satellite
diameter to primary diameter ratio is a little larger than most of the
ratios found for the largest TNOs.  There appears to be a difference
in the satellite to primary size ratios starting around the TNOs over
about 900 km.  Less than 900 km the satellite to size ratio increases
until near 400 km in primary size one starts to approach equal-sized
binaries.  For TNOs larger than 900 km, the satellite to primary size
ratio is generally smaller, with Pluto being an exception.

4) 2013 FY27 was monitored over minutes, hours and days with no
obvious short-term variability detected in the r-band.  The most
likely reason for no measureable short-term light curve is that 2013
FY27 is near spherical in shape with no significant albedo variations
on its surface to allow for a measurable rotational period.  2013 FY27
was further monitored in the r-band over months to find its linear
phase curve of $\beta = 0.25\pm 0.03$ mags/deg.  This is a slightly
higher than normal TNO phase curves, but reasonable for a moderate
albedo surface, unlike the lower phase curves found for very high
albedo TNO surfaces.  The reduced magnitude of 2013 FY27 was found to
be $m_{r}(1,1,0)=2.85\pm 0.02$ and $m_{V}(1,1,0)=3.15\pm 0.03$ mags.

\section*{Acknowledgments}
We thank D. Ragozzine for comments on the manuscript and sharing the
initial analysis of the 2013 FY27 satellite orbit.  This paper makes
use of the following ALMA data: ADS/JAO.ALMA\#2017.1.01662.S. ALMA is
a partnership of ESO (representing its member states), NSF (USA) and
NINS (Japan), together with NRC (Canada), NSC and ASIAA (Taiwan), and
KASI (Republic of Korea), in cooperation with the Republic of
Chile. The Joint ALMA Observatory is operated by ESO, AUI/NRAO and
NAOJ. The National Radio Astronomy Observatory is a facility of the
National Science Foundation operated under cooperative agreement by
Associated Universities, Inc.  This work is based in part on NASA/ESA
Hubble Space Telescope Cycle 25 Program 15248 observations.  This
paper includes data gathered with the 6.5 meter Magellan Telescopes
located at Las Campanas Observatory, Chile.

\newpage

%
%
%
%

\begin{center}
\begin{deluxetable}{llcccc}
\tablenum{1}
\tablewidth{6 in}
\tablecaption{Geometry of the Observations}
\tablecolumns{6}
\tablehead{
\colhead{Name} & \colhead{UT Date} & \colhead{$R$}  & \colhead{$\Delta$} & \colhead{$\alpha$} & \colhead{Telescope} \\ \colhead{} &\colhead{} &\colhead{(AU)} &\colhead{(AU)} & \colhead{(deg)} & \colhead{}}  
\startdata
2013 FY27        &  2016 Mar 08     &  80.26  &  79.30  &  0.17 & Magellan \nl
2013 FY27        &  2016 Mar 09     &  80.26  &  79.30  &  0.18 & Magellan \nl
2013 FY27        &  2016 Mar 10     &  80.26  &  79.30  &  0.18 & Magellan \nl
2013 FY27        &  2016 May 03     &  80.24  &  79.73  &  0.62 & Magellan \nl
2013 FY27        &  2017 Dec 29     &  80.09  &  79.74  &  0.66 & ALMA \nl
2013 FY27        &  2017 Dec 30     &  80.09  &  79.72  &  0.66 & ALMA \nl
2013 FY27        &  2018 Jan 04     &  80.09  &  79.64  &  0.63 & ALMA \nl
2013 FY27        &  2018 Jan 15     &  80.08  &  79.48  &  0.56 & HST \nl
\enddata
\tablenotetext{}{Quantities are the heliocentric distance ($R$), geocentric distance ($\Delta$) and phase angle ($\alpha$). UT Date shows the year, month and day of the observations.}
\end{deluxetable}
\end{center}

\newpage

%
%
%
%

\begin{center}
\begin{deluxetable}{lccc}
\tablenum{2}
\tablewidth{6.5 in}
\tablecaption{Optical Observations 2013 FY27 \label{tab:obskbo}}
\tablecolumns{4}
\tablehead{
\colhead{UT Date\tablenotemark{a}} & \colhead{Airmass} & \colhead{Exp\tablenotemark{b}}  & \colhead{Mag.\tablenotemark{c}}  \\  \colhead{yyyy mm dd hh:mm:ss}  & \colhead{} & \colhead{(sec)} & \colhead{($m_{r}$)} }
\startdata
 2016 03 08 01:43:32  &  1.36 &  225 &  21.91  \nl
 2016 03 08 01:48:45  &  1.34 &  225 &  22.00  \nl
 2016 03 08 02:20:38  &  1.23 &  225 &  21.96  \nl
 2016 03 08 02:25:53  &  1.21 &  225 &  21.99  \nl
 2016 03 08 02:56:24  &  1.15 &  225 &  21.95  \nl
 2016 03 08 03:25:11  &  1.10 &  300 &  21.94  \nl
 2016 03 08 04:00:14  &  1.08 &  250 &  21.91  \nl
 2016 03 08 04:46:55  &  1.08 &  250 &  21.89  \nl
 2016 03 08 05:36:42  &  1.13 &  300 &  21.95  \nl
 2016 03 08 07:14:42  &  1.45 &  350 &  21.95  \nl
 2016 03 08 07:35:49  &  1.58 &  350 &  21.95  \nl
 2016 03 09 00:30:34  &  1.86 &  400 &  21.95  \nl
 2016 03 09 00:39:18  &  1.77 &  400 &  21.98  \nl
 2016 03 09 01:15:27  &  1.49 &  400 &  21.99  \nl
 2016 03 09 01:23:45  &  1.44 &  380 &  21.98  \nl
 2016 03 09 01:31:36  &  1.40 &  380 &  21.92  \nl
 2016 03 09 01:54:21  &  1.30 &  400 &  21.92  \nl
 2016 03 09 02:02:28  &  1.27 &  400 &  21.92  \nl
 2016 03 09 02:24:01  &  1.21 &  380 &  21.92  \nl
 2016 03 09 02:31:47  &  1.19 &  400 &  21.96  \nl
 2016 03 09 03:04:00  &  1.13 &  400 &  21.94  \nl
 2016 03 09 03:12:11  &  1.12 &  350 &  21.96  \nl
 2016 03 09 04:11:13  &  1.07 &  330 &  21.92  \nl
 2016 03 09 04:55:11  &  1.09 &  350 &  21.93  \nl
 2016 03 09 05:54:18  &  1.17 &  350 &  21.93  \nl
 2016 03 09 07:21:05  &  1.51 &  330 &  21.94  \nl
 2016 03 09 07:38:33  &  1.63 &  280 &  21.98  \nl
 2016 03 10 01:01:19  &  1.55 &  330 &  21.93  \nl
 2016 03 10 01:31:04  &  1.38 &  330 &  21.93  \nl
 2016 03 10 02:19:26  &  1.21 &  300 &  21.92  \nl
 2016 03 10 02:48:42  &  1.15 &  250 &  21.89  \nl
 2016 03 10 03:31:37  &  1.09 &  275 &  21.92  \nl
 2016 03 10 05:08:40  &  1.11 &  220 &  21.90  \nl
 2016 03 10 06:31:05  &  1.29 &  280 &  21.90  \nl
 2016 03 10 07:38:46  &  1.67 &  250 &  21.94  \nl
 2016 05 02 23:28:15  &  1.13 &  300 &  22.06  \nl
 2016 05 02 23:33:56  &  1.12 &  250 &  22.04  \nl
 2016 05 03 00:47:00  &  1.08 &  200 &  22.03  \nl
 2016 05 03 02:14:34  &  1.17 &  250 &  22.06  \nl
 \enddata
\tablenotetext{a}{Universal date at the start of the integration.}
\tablenotetext{b}{Exposure time for the image.}
\tablenotetext{c}{Apparent red magnitude (r-band), uncertainties are $\pm 0.01$ to $0.02$.}
\end{deluxetable}
\end{center}

\newpage



\begin{center}
\begin{deluxetable}{ll}
\tablenum{3}
\tablewidth{4 in}
\tablecaption{Properties Found for 2013 FY27}
\tablecolumns{2}
\tablehead{
\colhead{Qtty}  & \colhead{Measurement} }  
\startdata
H       &  $3.15\pm 0.03$ mag   \nl
$m_{r}(1,1,0)$         &  $2.85\pm 0.02$ mags \nl
$m_{R}(1,1,0)$         &  $2.59\pm 0.02$ mags \nl
$m_{V}(1,1,0)$         &  $3.15\pm 0.03$ mags \nl
$m_{r}(\alpha = 0.18$ deg)     &  $21.92\pm 0.02$ mag  \nl
$m_{r}(\alpha = 0.62$ deg)     &  $22.04\pm 0.02$ mag  \nl
$g-r$     &  $0.76\pm 0.02$ mag  \nl
$r-i$&  $0.31\pm 0.03$ mag  \nl
$g-i$     &  $1.07\pm 0.04$ mag \nl
$m_{R}(\alpha = 0.62$ deg)     &      $21.79\pm 0.02$ mag \nl
$m_{V}(\alpha = 0.62$ deg)     &      $22.35\pm 0.03$ mag \nl
V-R     &  $0.56\pm 0.03$ mag \nl
R-I     &  $0.52\pm 0.03$ mag \nl
$S$     &  $17\pm 2$  \nl
$\beta$ &  $0.25 \pm 0.03$ \nl
$p_{v}$  & $0.17_{-0.030}^{+0.045}$ \nl
Effective Diameter  &  $765_{-85}^{+80}$ km \nl
Primary Diameter  &  $740_{-90}^{+85}$ km \nl
Satellite Diameter  &  $\sim 190$ km \nl
\enddata
\tablecomments{H is the absolute magnitude in the V-band.  The colors are from the Sloan g, r and i-bands and have also been converted to the Johnson-Kron-Cousins system B, V, R and I-bands using Smith et al. (2002).  $S$ is the spectral gradient of the object as described in  Sheppard (2010).}
\end{deluxetable}
\end{center}


\newpage

\newpage

\newpage

\newpage

\newpage

\newpage

\newpage

\newpage

\newpage

\newpage

\begin{references}


\reference{Bar05} Barucci, M., Belskaya, I., Fulchignoni, M. \& Birlan, M. 2005, AJ, 130, 1291.

\reference{Bar16} Barr, A. \& Schwamb, M. 2016, MNRAS, 460, 1542.

\reference{Bel00} Belskaya, I. \& Shevchenko, V. 2000, Icarus, 147, 94.

\reference{Bel08} Belskaya, I., Levasseur-Regourd, A., Shkuratov, Y. \& Muinonen, K. 2008, in The Solar System Beyond Neptune, ed. M. Barucci, H. Boehnhardt, D. Cruikshank and A. Morbidelli (Tucson: Univ of Arizona Press), 115-127.

\reference{Ben13} Benecchi, S. \& Sheppard, S. 2013, AJ, 145, 124.

\reference{Bro06} Brown, M., van Dam, M., Bouchez, A., et al. 2006, ApJ, 639, 43.

\reference{Bro12} Brown, M., Schaller, E. \& Fraser, W. 2012, AJ, 143, 146.

\reference{Bro13} Brown, M. 2013, ApJ, 778, L34.

\reference{Bro17} Brown, M. \& Butler, B. 2017, AJ, 154, 19.

\reference{Bro18} Brown, M. \& Butler, B. 2018, arXiv:1801.07221.

\reference{Bui97} Buie, M., Tholen, D. \& Wasserman, L. 1997, Icarus, 125, 233.

\reference{Can11} Canup, R., 2011, ApJ, 141, 35.

\reference{Dal15} Dalle Ore, C., Barucci, M., Emery, J. et al. 2015, Icarus, 252, 311.

\reference{Dor08} Doressoundiram, A., Boehnhardt, H., Tegler, S. and Trujillo, C. 2008, in The Solar System Beyond Neptune, ed. M. Barucci, H. Boehnhardt, D. Cruikshank and A. Morbidelli (Tucson: Univ of Arizona Press), 91-104.

\reference{Fer13} Fernandez, Y., Kelley, M., Lamy, P., et al. 2013, Icarus, 226, 1138.

\reference{Ferr16} Ferrari, C. \& Lucas, A. 2016, A\&A, 588, 133.

\reference{Fra15} Fraser, W., Brown, M. \& Glass, F. 2015, ApJ, 804, 31.

\reference{Ger17} Gerdes, D., Sako, M., Hamilton, S., et al. 2017, ApJ, 839, L15.

\reference{Gru09} Grundy, W. 2009, Icarus, 199, 560.

\reference{Gru15} Grundy, W., Porter, S., Benecchi, S. et al. 2015, Icarus, 257, 130.

\reference{Har98} Harris,  A. 1998, Icarus, 131, 291.

\reference{Hai12} Hainaut, O., Boehnhardt, H. \& Protopapa, S. 2012, A\&A, 546, 115.

\reference{Jew02} Jewitt, D. 2002, AJ, 123, 1039.

\reference{Kis17} Kiss, C., Marton, G., Farkas-Takacs, A. et al. 2017, ApJ, 838, 1.

\reference{Kov17} Kovalenko, I., Doressoundiram, A., Lellouch, E., Vilenius, E., Muller, T. \& Stansberry, J. 2017, A\&A, 608, 19.

\reference{Lac14} Lacerda, P., Fornasier, S., Lellouch, E., et al. 2014, ApJ, 793, L2.

\reference{Leb89} Lebofski et al. 1989, Icarus, 78, 335

\reference{Lel13} Lellouch, E., Santos-Sanz, P., Lacerda, P. et al. 2013, A\&A, 557, A60.

\reference{Lel17} Lellouch, E., Moreno, R., Muller, T., et al. 2017, A\&A, 608, 45.

\reference{Mac07} McMullin, J. P., Waters, B., Schiebel, D., Young, W., \& Golap, K. 2007, Astronomical Data Analysis Software and Systems XVI (ASP Conf. Ser. 376), ed. R. A. Shaw, F. Hill, \& D. J. Bell, 127.

\reference{Mac14} MacKenty, J., Baggett, S., Brammer, G., Hilbert, B.,
Long, K., McCullough, P. \& Adam, G. 2014, Proc. SPIE, 9143, 914328.

\reference{McK17} McKinnon, W., Stern, S., Weaver, H., et al. 2017, Icarus, 287, 2.

\reference{Nes10} Nesvorny, D., Youdin, A., \& Richardson, D. 2010, AJ, 140, 785.

\reference{Nol08} Noll, K., Grundy, W., Chiang, E., Margot, J. \& Kern, S. 2008, in The Solar System Beyond Neptune, ed. M. Barucci, H. Boehnhardt, D. Cruikshank and A. Morbidelli (Tucson: Univ of Arizona Press), 345-363.

\reference{Par11} Parker, A., Kavelaars, J., Petit, J., Jones, L., Gladman, B. \& Parker, J. 2011, ApJ, 743, 1.

\reference{Par16} Parker, A., Buie, M., Grundy, W., \& Noll, K. 2016, ApJ, 825, 9.

\reference{Rab07} Rabinowitz, D., Schaefer, B. \& Tourtellotte, S. 2007, AJ, 133, 26.

\reference{San17} Santos-Sanz, P., Lellouch, E., Groussin, O., Lacerda, P., Muller, T., Ortiz, J., Kiss, C., Vilenius, E., Stansberry, J., Duffard, R., Fornasier, S., Jorda, L. \& Thirouin, A. 2017, A\&A, 604, 95.

\reference{Sch08} Schlichting, H. \& Re'em, S. 2008, ApJ, 673, 1218.

\reference{She02} Sheppard, S. \& Jewitt, D. 2002, AJ, 124, 1757.

\reference{She03} Sheppard, S. \& Jewitt, D. 2003, EM\&P, 92, 207.

\reference{She07} Sheppard, S. 2007, AJ, 134, 787.

\reference{She10} Sheppard, S. 2010, AJ, 139, 1394.

\reference{She12} Sheppard, S. 2012, AJ, 144, 169.

\reference{She12} Sheppard, S., Ragozzine, D., \& Trujillo, C. 2012, AJ, 143, 58.

\reference{She16} Sheppard, S. \& Trujillo, C. 2016, AJ, 152, 221.

\reference{She18} Sheppard, S. 2018, CBET, 4537.

\reference{Smi02} Smith, J., Tucker, D., Kent, S., et al. 2002, AJ, 123, 2121.

\reference{Sta08} Stansberry, J., Grundy, W., Brown, M., Cruikshank, D., Spencer, J., Trilling, D. and Margot, J. 2008, in The Solar System Beyond Neptune, ed. M. Barucci, H. Boehnhardt, D. Cruikshank and A. Morbidelli (Tucson: Univ of Arizona Press), 161-179.




\end{references}
\end{document}